\documentclass[prc,showpacs,twocolumn]{revtex4}

\usepackage{graphicx}
\def\pt{p_T}
\def\xb{\bar\xi}
\def\dis{distribution}
\def\pb{\phi,b)}

\begin{document} 

\title
 {Scaling Behavior of the Azimuthal and Centrality Dependencies of Jet Production in Heavy-ion Collisions}
\author
 {Rudolph C. Hwa$^1$ and C.\ B.\ Yang$^{1,2}$}
\affiliation
{$^1$Institute of Theoretical Science and Department of
Physics\\ University of Oregon, Eugene, OR 97403-5203, USA\\
$^2$Institute of Particle Physics, Hua-Zhong Normal
University, Wuhan 430079, P.\ R.\ China}

\begin{abstract} 
For heavy-ion collisions at RHIC  a scaling behavior is found in the 
dependencies on azimuthal angle $\phi$ and impact parameter $b$ for pion production at high $\pt$ essentially independent of the hadronization process. The scaling variable is in terms of a
dynamical path length $\xi$ that takes into account detailed properties of geometry, medium density and probability of hard scattering. It is shown in the recombination model how the nuclear modification factor  depends on the average $\bar\xi(\phi,b)$. The data for $\pi^0$ production at $\pt=$ 4-5 and 7-8 GeV/c at RHIC are shown to exhibit the same scaling behavior as found in the model calculation. Extension to back-to-back dijet production has been carried out, showing the existence of $\bar\xi$ scaling also in the away-side yield per trigger. At LHC the hard-parton density can be high enough to realize the likelihood of recombination of shower partons arising from neighboring jets. It is shown that such 2-jet recombination can cause strong violation of $\bar\xi$ scaling. Furthermore, the large value of $R_{AA}$ that exceeds 1 can become a striking signature of such a hadronization process at high energy.
\end{abstract}
\pacs{25.75.-q, 25.75.Gz, 24.85.+p}
\maketitle
\section{Introduction}

Recent analysis of the high-statistics data on $\pi^0$ production in heavy-ion collisions has provided a wealth of information on the nuclear modification factor ($R_{AA}$) as a function of centrality ($c$), azimuthal angle ($\phi$), and transverse momentum ($p_T$) \cite{af}.  The complexity revealed is clearly in need of some organizational simplification, if useful insight is to be gained from it to advance a theoretical understanding of the various features.  For example, the $p_T$ measured ranges from 1 to 10 GeV/c; the conventional wisdom is to separate the high and low regions and treat them with different dynamics.  The $\phi$ dependence is studied by determining the second harmonics ($v_2$), whose dependencies on $c$ and $p_T$ are presented as if they are variables orthogonal to $\phi$, although it is known that $c$ and $\phi$ are correlated by geometry and $p_T$ depends on path length.  On the theory side treatments of fundamental issues at the microscopic level cannot incorporate experimental details without sacrificing clarity and precision.  In the interest of finding simplifying features of the data we examine in detail the geometrical aspect of the problem associated with the propagation of a hard parton through the medium and explore the possibility of scaling behavior in a variable, $\bar\xi$, that involves both impact parameter ($b$) and $\phi$.  On the basis of that result we then organize the dynamical problem of hadronization in a way that can exhibit the scaling behavior.  With that behavior derived on a theoretical basis in certain approximation, the data on $R_{AA}$ can then be shown in the same format to check the relevance of the model-dependent finding.  A form of universality does exist in reality.

At low $p_T$ where thermal and semihard partons dominate, a study of the surface factor $S(\phi,b)$ that relates $b$ and $\phi$ has led to the elucidation of the $\phi$ dependencies of $R_{AA}(\phi,b)$ and ridge yield $Y^R(\phi,b)$ \cite{hz}.  At higher $p_T$ shower partons become more important \cite{hy1}.  Accordingly, the path length of a hard parton in the dense medium becomes the major factor that affects $R_{AA}$.  That length depends on $\phi, b$, and the point of creation of the hard parton, the probability of which depends on the nuclear overlap function.  All those quantities are calculable from geometrical considerations.  Of course, at some point dynamics enters the picture.  Recombination can handle well the late-stage hadronization process at all $p_T$ \cite{hy1}.  Our primary objective is to expose the simplifying features of the nuclear complexity that a hard parton experiences at the early stage of the production process. It is complementary to the study of the jet quenching process in QCD that assumes a simple path length $L$ \cite{rd,dk}. The two meet at the point where momentum degradation is expressed in terms of the scaling variable $\bar\xi(\phi,b)$. The connection is complicated by the role that thermal partons play in the hadronization of shower partons at the intermediate $\pt$ range. Nevertheless, a simple behavior of $R_{AA}$ in terms of $\bar\xi$ can be determined theoretically and experimentally.

If indeed a universal behavior of $R_{AA}$ can be found in its dependence on a dynamical path length, then it should be natural to follow up the investigation with a study of the properties of back-to-back dijets, since the dependence of the yield on the momenta of the trigger and away-side particles ($p_t$ and $p_b$, respectively) is closely related to the path lengths traveled by the hard parton and its recoil.  A detailed study of the dependence on $p_t$ and $p_b$ has been carried out earlier, but with $\phi$ averaged over \cite{hy2}.  Now, with specific $\phi$ values taken into account, the joint dependence on $b$ and $\phi$ presents complications that must be simplified, if useful comparison between theory and experiment is to be facilitated.

Dijets mentioned above are directed in opposite directions, but two jets going in the same direction would be of tremendous importance when collision energy is very high as at LHC. Pion production arising from 2-jet recombination should overwhelm 1-jet fragmentation when the rate of hard scattering is significantly increased at high initial energy. Our procedure of studying the $\phi$ and $b$ dependence can readily be extended to the coalescence of shower partons from parallel jets. Our result shows the possibility of dramatic departure from the usual expectation on the nuclear modification factor.

In this work we shall focus on the production of pions, since the high-statistics data \cite{af} that can check our analysis are for $\pi^0$ only. The extension to the production of protons and other hadrons can be done in the same framework, but will not be carried out here.

\section{Inclusive pion DISTRIBUTION AT FIXED AZIMUTHAL ANGLE}

We shall work in a $p_T$ range where thermal-thermal recombination is a small correction to what we shall calculate.  That would be for $p_T\ ^>_\sim\ 4$ GeV/c at $\sqrt{s}=200$ GeV.  It does not mean that thermal partons are unimportant because their recombination with shower partons (the TS component) can be dominant (for $\pi$ production).  Both TS and SS components are proportional to the probability of hard-parton production, so we may write the basic structure of the single-particle distribution in nuclear collisions in the form
\begin{eqnarray}
&&{dN_{AA}\over p_Tdp_Td\phi}(b)\equiv \rho_1(\pt,\phi,b) \nonumber \\ 
&=&\int{dq\over q}\sum_i F_i(q,\phi,b) H_i(q,\pt) + \rho_1^{\rm TT}(\pt,\phi,b),  \label{1}
\end{eqnarray}
where $F_i (q, \phi, b)$ is the probability that a parton of species $i$ with momentum $q$ at azimulthal angle $\phi$ (relative to the reaction plane, or more precisely relative to the minor axis of the overlap almond in our theoretical consideration) emerges at the surface of medium.  We consider only the transverse plane at mid-rapidity.  $H_i(q, p_T)$ describes the hadronization of parton $i$ with momentum $q$ to a hadron of whatever type under consideration with momentum $p_T$.  In the recombination model the medium effect on hadronization is taken into account by the  TS component in $H_i$ in addition to the fragmentation term that is represented by the SS component \cite{hy1, hy3}.  Such details are not relevant to our concern here.  The main point in Eq.\ (\ref{1}) is that the dependence on $\phi$ and $b$ are in $F_i(q, \phi, b)$.   The thermal partons can have a minor dependence of those variables, and is an issue that will be discussed later in this paper.

The essence of our problem is the effect of the medium on the hard parton from creation to its emergence on the surface.  For every hard parton created with momentum $k$ the probability of its emergence with momentum $q$ may be denoted by $J(k, q, \phi, b)$ so that we have
\begin{eqnarray}
F_i(q,\phi,b)=\int dk k f_i(k) J(k,q,\phi,b) ,  \label{2}
\end{eqnarray}
where $f_i(k)$ is the distribution for parton momentum $k$ at the creation point and has been parametrized in Ref.\ \cite{sg}.  The momentum degradation factor $J(k, q, \phi, b)$ strictly should depend on the parton type, but we make the approximation of ignoring that dependence, since our focus is on the path length dependence.  To show explicitly how path length enters the problem, we must exhibit the creation point at $(x_0, y_0)$, which is to be integrated over the region of nuclear overlap, weighted by the probability of a hard collision.  That is, we write
\begin{eqnarray}
J(k,q,\phi,b)=\int dx_0 dy_0 Q(x_0,y_0,b) G(k,q,\ell\left(x_0,y_0,\phi,b)\right),  \label{3}
\end{eqnarray}
where $Q(x_0, y_0, b)$ is the overlap function to be detailed below, and $G(k, q, \ell)$ is the momentum degradation factor that changes the parton momentum from $k$ to $q$ in a length $\ell$.  That length depends on the creation point $(x_0, y_0)$, the angle $\phi$ that the hard parton is directed, and the exit point $(x_1, y_1)$ for a nuclear collision at impact parameter $b$.  Again, we defer the calculation of $\ell(x_0, y_0, \phi, b)$ until later so that our description of the general scheme is not interrupted.

Putting Eqs.\ (\ref{2}) and (\ref{3}) in (1), we can break up the long equation into four parts by introducing $\int d\xi\delta(\xi-\gamma\ell)$, and write
\begin {eqnarray}
P(\xi,\phi,b)&=&\int dx_0dy_0 Q(x_0,y_0,b) \nonumber \\ 
&&\quad\times\delta (\xi-\gamma\ell(x_0, y_0, \phi, b)),  \label{4}\\
F_i(q,\phi,b)&=&\int d\xi P(\xi,\phi,b) F_i(q,\xi) , \label{5} \\
F_i(q,\xi)&=&\int dk k f_i(k) G(k,q,\xi) ,  \label{6}\\
\rho_1^{\rm TS+SS}(\pt,\phi,b)&=&\int {dq\over q}\sum_i F_i(q,\phi,b) H_i(q,p_T) . \label{7}
\end{eqnarray}
A parameter $\gamma$ is introduced in Eq.\ (\ref{4}) to represent the dynamical effect of energy loss on the variable $\xi$ that we call the dynamical path length.  If there is no energy loss, then the medium is transparent and $\xi = 0$ whatever the kinematical path length $\ell$ may be.  The value of $\gamma$ will be determined by phenomenology later, after the form of momentum degradation, specified by $G(k, q, \xi)$, is discussed.  The equations from (\ref{4}) to (\ref{7}) exhibit the different parts of the production process separately:  (\ref{4}) shows the probability of having a dynamical path length $\xi$ for a parton directed at $\phi$; (\ref{5}) gives the distribution of a parton with momentum $q$ at the surface at angle $\phi$; (\ref{6}) is the corresponding distribution after traversing an absorptive distance $\xi$ in the medium; (\ref{7}) is the final hadron distribution at $p_T$ and $\phi$.  A point to note is that the relationship between $\phi$ and $b$ that depends on the initial configuration is completely contained in Eq.\ (\ref{4}).

We now define the quantities introduced earlier in Eq.\ (\ref{4}) in the framework of the Glauber model for $AB$ collisions.  The thickness function $T_A(s)$ normalized to $A$ is
\begin{eqnarray}
T_A(s)=A\int dz \rho(s,z) ,  \qquad  \int d^2s T_A(s)=A ,  \label{8}
\end{eqnarray}
where $\rho$ is the nuclear density normalized to 1
\begin{eqnarray}
\rho(r)=\rho_0 [1+e^{(r-r_0)/\delta}]^{-1}  \label{9}
\end{eqnarray}
with $r_0=6.45$ fm and $\delta=0.55$ fm for Au.  We shall use $R_A=7$ fm for the effective nuclear radius.  For any point $(x, y)$ in the transverse plane, we have 
\begin{eqnarray}
s^2&=&(x+b/2)^2+y^2 ,  \label{10}  \\
z_A^2&=&1-s^2,   \qquad  z_B^2=1-|\vec s-\vec b|^2 ,  \label{11}
\end{eqnarray}
where all lengths are scaled to $R_A$.  The longitudinal lengths of $A$ and $B$ at $(x, y)$ are
\begin{eqnarray}
L_{A,B}(x,y) = {1\over \rho_0} \int_{-z_{A,B}}^{z_{A,B}} dz\ \rho(s,z) , \label{12}
\end{eqnarray}
where $\rho_0=0.285$.  With $\sigma$ being the inelastic nucleon-nucleon cross section, $T_A(s)$ and $L_A(x, y)$ are related by \cite{ch}
\begin{eqnarray}
\sigma T_A(s)=\omega L_A(x,y) ,  \qquad \omega=\sigma A\rho_0=4.6 \label{13}
\end{eqnarray}
for $A=197$.  The probability for producing a hard parton at $(x_0, y_0)$ is proportional to $T_A(\vec s+\vec b/2)T_B(\vec s-\vec b/2)$, where $(x_0, y_0)$ are the Cartesian coordinates of $\vec s$, so the normalized $Q(x_0, y_0, b)$ is
\begin{eqnarray}
Q(x_0,y_0,b)={T_A(x_0,y_0,-b/2)\, T_B(x_0,y_0,b/2)\over \int d^2s T_A(\vec s+\vec b/2) T_B(\vec s-\vec b/2)} ,  \label{14}
\end{eqnarray}
where $T_B(x_0, y_0, b/2)$ is the thickness function of nucleus $B$ whose center is at an impact parameter $b$ on the $x$ axis from that of $A$, located at $x=-b/2$.  

The path length from the point $(x_0, y_0)$ of creation to the exit point $(x_1, y_1)$ on the boundary is
\begin{eqnarray}
\ell(x_0,y_0,\phi,b)=\int_0^{t_1(x_0,y_0,\phi,b)} dt\, D(x(t),y(t)) ,  \label{15}
\end{eqnarray}
where the integrand is weighted by the local density along the trajectory marked by $t$, since energy loss is proportional to the medium density, and our aim here is to calculate the geometrical part of the factors that contribute to momentum degradation.  Apart from an overall normalizaation constant, that density inside the almond-shaped overlap region is given by
\begin{eqnarray}
D_a(x,y)&=&\omega L_A(x,y)[1-e^{-\omega L_B(x,y)}] \nonumber\\
&+&\omega L_B(x,y)[1-e^{-\omega L_A(x,y)}] .  \label{16}
\end{eqnarray}
The normalization is not of concern here because $\ell(x_0, y_0, \phi, b)$ appears only in conjunction with $\gamma$ in Eq.\ (\ref{4}), where $\gamma$ is an adjustable parameter to be fixed later in fitting the data.

The exit point $(x_1, y_1)$ can be determined in terms of $(x_0, y_0)$ and $\phi$.  The initial point is inside the overlap region, where $Q(x_0, y_0, b)$ is non-vanishing.  The exit point is somewhere on the elliptical boundary, since some time elapses before the hard parton reaches the surface.  There is no rigorous way to calculate the transition from the almond region to the ellipse, since hydrodynamics is not valid before local equilibrium is established; besides, semihard partons and the ridge formation that they give rise to can significantly change the boundary, unaccounted for by collective flow \cite{hz, chy, ch}.  We determine $(x_1, y_1)$ and the local density $D(x, y)$ along the trajectory in the following way.  First, we take the boundary of the ellipse to satisfy
\begin{eqnarray}
\left({x\over w} \right)^2+\left({y\over h} \right)^2=1 ,  \label{17}
\end{eqnarray}
where
\begin{eqnarray}
w=1-b/2 ,  \qquad  h=(1-b^2/4)^{1/2} ,  \label{18}
\end{eqnarray}
independent of the transit time from $(x_0, y_0)$ to $(x_1, y_1)$.  We map the density $D_a(x, y)$ for the almond region to the density $D(x, y)$ of the elliptical region by the identification
\begin{eqnarray}
D(x,y) = D_a(xd_1/d_2, yd_1/d_2)  \label{19}
\end{eqnarray}
along any radial direction measured from the origin where $x=y=0$.  The distances from the origin to the almond and elliptical boundaries are denoted by $d_1$ and $d_2$, respectively, and described in the Appendix.  
The ratio $d_1/d_2$ is very nearly 1 for almost all angles, since $w$ and $h$ are the $x$- and $y$-axis intercepts of both boundaries. We take the trouble to do the mapping in Eq.\ (\ref{19}) for the short duration before thermal equilibrium is established in order to render Eq.\ (\ref{15}) well defined, where the upper limit is at the boundary of the ellipse.
If the point $(x_0,y_0)$ is far from the boundary, the transit time for the hard parton to arrive at the medium boundary can be large so the ellipse would be larger than that described by Eq.\ (\ref{17}). 
For large transit time the dynamics of medium expansion predominantly in the longitudinal direction should be taken into consideration.
In that case we rely on the dynamical scaling behavior found in Ref.\ \cite{sw} to regard our procedure   as being insensitive to that expansion. Specifically, it means that if the RHS of Eq.\ (\ref{17}) were larger, $d_2$ would be larger, and the rescaling of $D(x,y)$  would redistribute the density  accordingly and the net result of the integration in Eq.\ (\ref{15}) would not deviate by too much.  $D_a(x_a, y_a)$ in Eq.\ (\ref{19}) is identified with the density given in (\ref{16}) for every point $(x_a, y_a)$ inside the almond.  
By restricting Eq.\ (\ref{15}) to only the short time duration when (\ref{19}) is valid,  it is much easier to calculate the exit point $(x_1, y_1)$, where the hard parton trajectory along $\phi$ intersects the ellipse, i.e., 
\begin{eqnarray}
x_1=x_0+t_1\cos\phi ,  \qquad y_1=y_0+t_1\sin \phi ,  \label{20}
\end{eqnarray}
where $t_1(x_0, y_0, \phi, b)$ can be determined by solving Eq.\ (\ref{17}), as done in the Appendix.  That is what we use as the upper limit of the integration in Eq.\ (\ref{15}), thereby giving dependence on $x_0$, $y_0$, $\phi$ and $b$ to the length $\ell(x_0, y_0, \phi, b)$.  
The result gives an estimate of the path length that may possess more general validity than what the static approximation of ignoring explicit transverse expansion may imply.
In this way $P(\xi, \phi, b)$, as defined in Eq.\ (\ref{4}), is completely calculable from geometrical considerations.  

\section{SCALING BEHAVIOR}

We now focus on $P(\xi, \phi, b)$ and try to extract as much information as possible about its dependence on $\phi$ and $b$ before proceeding to the dynamical problem of momentum degradation and particle production.  From Eqs.\ (\ref{4}) and (\ref{14}) $P(\xi, \phi, b)$ is properly normalized as
\begin{eqnarray}
\int d\xi P(\xi,\phi,b)=1 ,  \label{21}
\end{eqnarray}
so the mean $\bar\xi$ is
\begin{eqnarray}
\bar\xi(\phi,b)=\int d\xi\, \xi P(\xi,\phi,b) .  \label{22}
\end{eqnarray}
Since centrality is more directly accessible in experiments, we shall freely change from the dependence on $b$ to that on $c$, which is our symbol for centrality; for example, $c=0.05$ stands for 0-10\% centrality.  The relationship between $b$ and $c$ is well established, and is tabulated in Ref.\ \cite{ab}, for instance.

Using Eq.\ (\ref{4}) we obtain
\begin{eqnarray}
\bar\xi(\phi,c(b))=\gamma\int dx_0dy_0 \ell(x_0,y_0,\phi,b) Q(x_0,y_0,b) .  \label{23}
\end{eqnarray}
In the next section we shall show that $\gamma=0.11$.  For our general discussion here $\gamma$ need not be specified, although the numerical value will be used when plots are presented in figures.  We show in Fig.\ 1 the dependence of $\bar\xi$ on $\phi$ for six values of $c$ for AuAu collisions at $\sqrt{s}=200$ GeV.  Except for a shift in the magnitude, that dependence seems to be insensitive to $c$ for $c$ ranging from 0.05 to 0.55, which corresponds to $b$ from 0.48 to 1.58 in units of $R_A$.  It suggests that there is a universality in the $\phi$ dependence for $c\ge 0.05$. That universality cannot be exact, since we know that when $b=0$ there can be no dependence on $\phi$. However, for $c>0.05$ approximate universality seems valid in Fig.\ 1.  If so, then the question becomes whether the probability distribution $P(\xi, \phi, b)$ that is a function of three variables can more economically be expressed in terms of few variables.

\begin{figure}[tbph]
\includegraphics[width=0.4\textwidth]{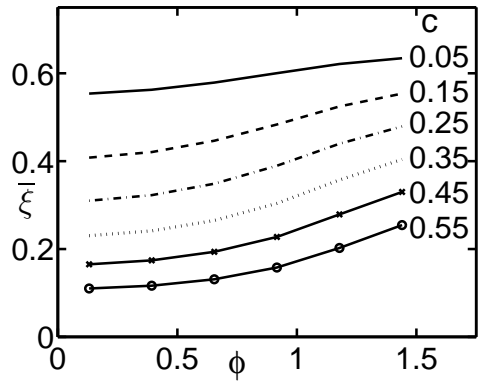}
\caption{Average dynamical path length $\bar\xi$ vs $\phi$ for six values of centrality.}
\end{figure}

Our first step in addressing that question is to ask whether, at fixed $b$, the \dis\ is a scaling function. More specifically, we define
\begin{eqnarray}
z=\xi/\xb ,  \label{24}
\end{eqnarray}
and ask whether $P(\xi,\phi,b)$ can be written in terms of a scaling function $\psi(z)$ in the form
\begin{eqnarray}
P(\xi,\phi,b)=\psi(z)/\xb(\phi,b)  \label{25}
\end{eqnarray}
such that
\begin{eqnarray}
\int dz\psi(z)=\int dz z\psi(z) =1 .  \label{26}
\end{eqnarray}
Distributions that have such properties are referred to as satisfying KNO scaling, well-known in multiparticle production \cite{kno,kew}. We show in Fig.\ 2 the behavior of $\psi(z)$ for (a) $c=0.05$ and (b) $c=0.55$; in each case there are six values of $\phi$: $\phi=n\pi/24, n=1,3,5,\cdots 11.$ We use $c$ for classification, instead of $b$, because the data to be shown in the next section are given in terms of centrality. What is notable is that in each case all dependencies on $\phi$ collapse to one universal curve in terms of $z$, when expressed in the format of Eq.\ (\ref{25}). That is scaling in $\phi$. The next question is whether there is scaling in $c$.

\begin{figure}[tbph]
\includegraphics[width=0.35\textwidth]{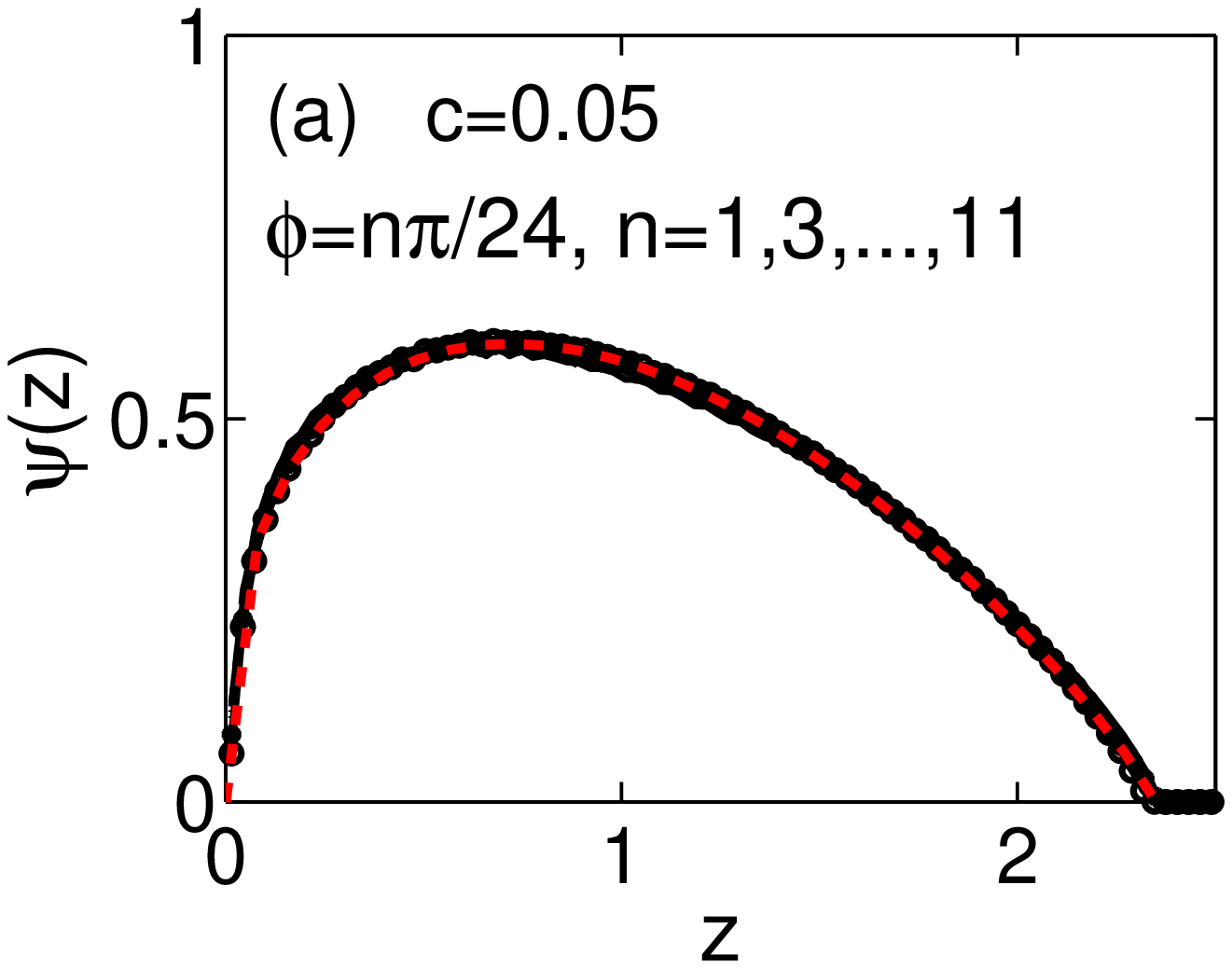}
\includegraphics[width=0.35\textwidth]{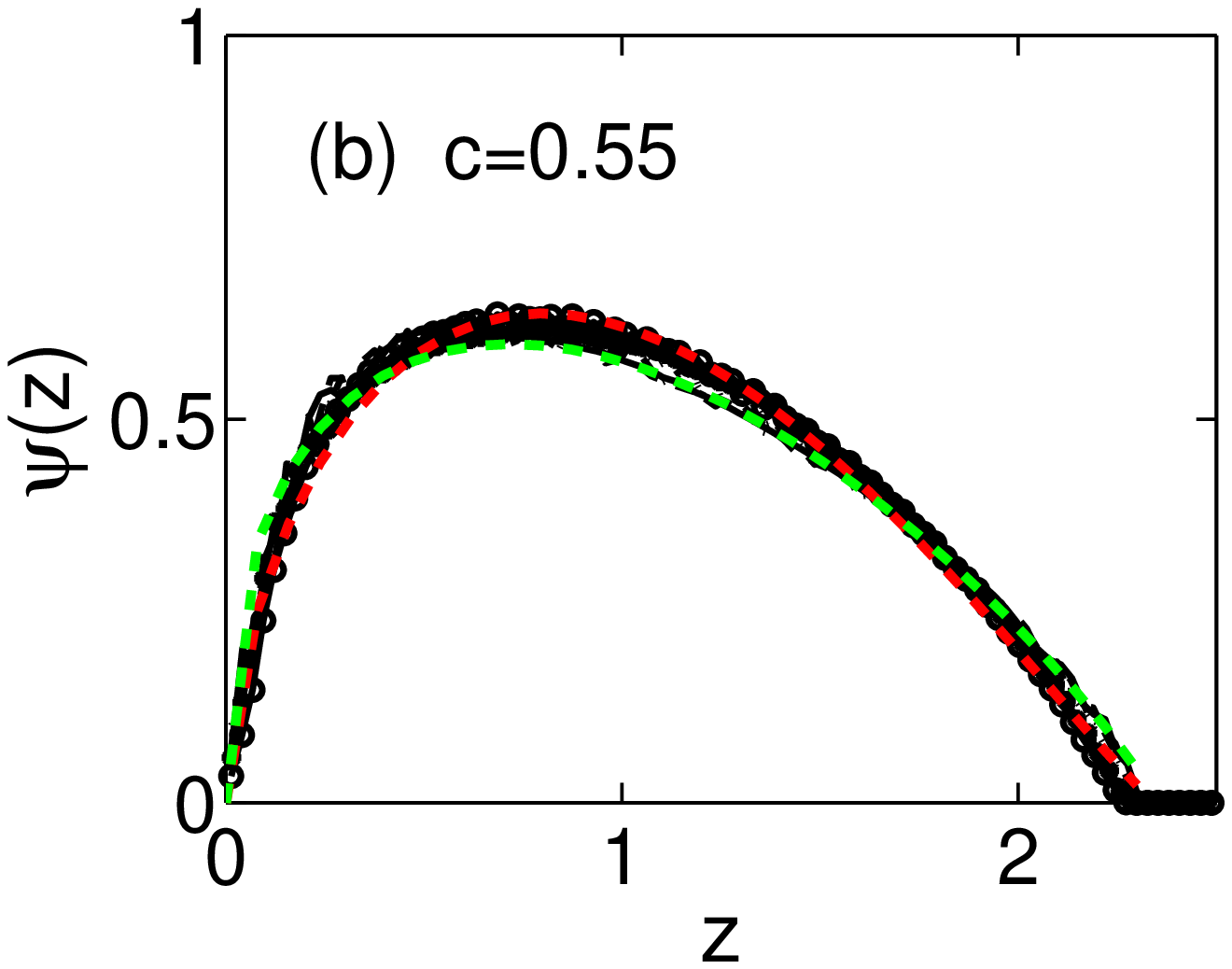}
\caption{(Color online) Scaling function $\psi(z)$ for six values of $\phi$ at (a) 0-10\% centrality and (b) 50-60\% centrality. The red lines are fits of the two figures, parametrized by Eq.\ (\ref{28}) and (\ref{29});  the green line in (b) is a reproduction of the red line in (a).}
\end{figure}

To quantify the dependence on $c$, we fit $\psi(z)$ by a 2-parameter formula
\begin{eqnarray}
\psi(z)= \zeta^{a_1}(1-\zeta)^{a_2}/B(a_1+1,a_2+1),  \  \zeta=z/2.4,  \label{27}
\end{eqnarray}
where $B(\alpha,\beta)$ is the Beta function. The best fits yield
\begin{eqnarray}
(a)\quad c=0.05, \quad a_1=0.37,  \quad a_2=0.81,   \label{28} \\
(b)\quad c=0.55, \quad a_1=0.57,  \quad a_2=1.05.  \label{29}
\end{eqnarray}
They are shown by the (red) dashed lines in Fig.\ 2(a) and (b). The (green) dashed line in Fig.\ 2(b) is a reproduction of the (red) dashed line in Fig.\ 2(a) for the purpose of comparison. Visually, the two scaling curves do not differ by too much, although numerically the values of $a_1$ and $a_2$ in Eqs.\ (\ref{28}) and (\ref{29}) are noticeably different between the two cases. To a 10\% accuracy, one may regard $\psi(z)$ to be universal for $c$ ranging from 0.05 to 0.55 and $\phi$ from 0 to $\pi/2$. That offers a remarkable degree of simplicity in the complex geometrical problem of nuclear collisions. In words it means that $P(\xi,\phi,c)$ depends only on $\xi$ and $\xb$, not on $\phi$ and $c$ separately.

\section{NUCLEAR MODIFICATION FACTOR}

Having found some general properties of the collision system that are basically geometrical, we now proceed to the problem of hadron production and see how the geometrical insight gained in the preceding section can help us organize the dependence on $\phi$ and $c$.  As a consequence we shall find an efficient way to compare our result with the data on $R_{AA}(p_T, \phi, c)$.  

Let us start by inserting some details that are omitted in Sec.\ 2.  Since the PHENIX data \cite{af} on $R_{AA}$ are for $\pi^0$ production, we review the formalism for calculating the single-pion inclusive distribution \cite{hy1}.  We restrict our consideration to midrapidity and  write the invariant distribution in the recombination model for a pion in the direction $\vec p$ in the 1-dimensional form as
\begin{eqnarray}
p^0{dN_{\pi}\over dp}=\int {dq_1\over q_1}{dq_2\over q_2} F_{q\bar q}(q_1,q_2) R_{\pi}(q_1,q_2,p) ,  \label{30}
\end{eqnarray}
where the $q\bar q$ \dis\ is in general
\begin{eqnarray}
F_{q\bar q}(q_1,q_2)={\cal TT+TS+SS} ,  \label{31}
\end{eqnarray}
and the recombination function (RF)
\begin{eqnarray}
R_{\pi}(q_1,q_2,p)={q_1q_2\over p^2} \delta\left({q_1\over p}+{q_2\over p}-1\right) . \label{32}
\end{eqnarray}
To reproduce the exponential behavior of the observed distribution at low $p_T$ by TT recombination, we have assumed the thermal parton distribution to have the form 
\begin{eqnarray}
{\cal T}(q_1)=q_1{dN_q^{\rm th}\over dq_1}=Cq_1e^{-q_1/T} ,  \label{33}
\end{eqnarray}
so that the thermal pion \dis\ is
\begin{eqnarray}
{dN_{\pi}^{\rm TT}\over p_Td\pt}={C^2\over 6} e^{-\pt/T} .  \label{34}
\end{eqnarray}
For the shower parton \dis\ we have
\begin{eqnarray}
{\cal S}(q_2)=\int {dq\over q}\sum_i F_i(q) S_i^j(q_2/q) ,  \label{35}
\end{eqnarray}
where $S_i^j(z_2)$ is the matrix of shower parton determined from the fragmentation functions \cite{hy3}, and $F_i(q)$ is the hard parton \dis\ at the medium surface before fragmentation. The TS contribution to the final pion \dis\ is
\begin{eqnarray}
{dN_{\pi}^{\rm TS}\over \pt d\pt}&=&{1\over \pt^2} \sum_i \int {dq\over q} F_i(q)\widehat{\sf TS}  (q,\pt) ,  \label{36}  \\
\widehat{\sf TS}  (q,\pt)&=&\int {dq_2\over q_2} S_i^j(q_2/q) \int dq_1 Ce^{-q_1/T} \nonumber \\
&& \qquad\times R(q_1,q_2,\pt) .  \label{37}
\end{eqnarray}
Finally, the SS component that is equivalent to the fragmentation component is
\begin{eqnarray}
{dN_{\pi}^{\rm SS}\over \pt d\pt}={1\over \pt^2}\sum_i \int {dq\over q}F_i(q){\pt\over q}D_i^{\pi}\left({\pt\over q}\right) ,  \label{38}
\end{eqnarray}
where $D(z)$ is the fragmentation function.
When these equations are compared to Eq.\ (\ref{1}), it should be clear what $H_i(q,\pt)$ in the latter stands for.

The above is a summary of the basic formulas when $\phi$ if averaged over.  Now, as we consider $\phi$ and $b$ dependencies explicitly, let us revisit the recombination process. A quark $q$ and an anitquark $\bar q$ with momenta $\vec q_1$ and $\vec q_2$ are unlikely to recombine to form a pion, if $\vec q_1$ and $\vec q_2$ are not approximately parallel, since their relative momentum normal to the pion momentum should not be larger than the inverse size of the pion.  Thus we shall continue to use the 1-dimensional description of the RF.  However, the $\phi$ distributions of the coalescing partons should be taken into account.  

For the thermal partons a description is given in Ref.\ \cite{hz} that can simultaneously reproduce three empirical facts, which are:  (a) $R_{AA}(p_T, \phi, c)$ at $1<p_T<2$ GeV/c \cite{af}, (b) $v_2(p_T, c)$ \cite{af}, and (c) ridge yield $Y^R(\phi_s, c)$ \cite{fen}.  The physics developed there involves a consideration of ridge formation with or without triggers.  We omit a description of that here, since the details are not of crucial importance to our calculation below.  We simply adopt the thermal parton distribution in the form  
\begin{eqnarray}
{\cal T}(q_1,\phi,b)=C(b)q_1[1+{1\over 2}aD(x',y',b)S(\pb]e^{-q_1/T} ,  \label{39}
\end{eqnarray}
where a small $\phi$-dependent term is included in the square brackets that represents the effect of semihard scattering. $D(x',y',b)$ is as defined in Eq.\ (\ref{16}) with $(x',y')$ being evaluated at 0.17 from the boundary of the almond overlap on the $x$ axis. $S(\pb$ is a surface factor defined by
\begin{eqnarray}
S(\pb=h[E(\theta_2,\alpha)-E(\theta_1,\alpha)] ,  \label{40}
\end{eqnarray}
where $E(\theta,\alpha)$ is the elliptic integral of the second kind, $\theta_i=\tan^{-1}({h\over w}\tan\phi_i), \phi_1=\phi-\sigma, \phi_2=\phi+\sigma, \sigma=0.33,$ and $\alpha(b)=1-w^2/h^2$. It causes a small enhancement of the thermal partons due to semihard partons with $T=0.317$ GeV and $a=0.47$ in Eq.\ (\ref{39}). $C(b)$ is given in terms of $N_{\rm part}$ as
 \begin{eqnarray}
C(N_{\rm part})=1.1 N_{\rm part}^{0.54}\  {\rm GeV}^{-1} .  \label{41}
\end{eqnarray}
These complications are included here for completeness, even though their effect on the final result is small.

The shower partons, on the other hand, depend crucially on $\phi$, as a comparison between Eqs.\ (\ref{7}) and (\ref{35}) reveals the presence of $\phi$ in $F_i(q, \phi, b)$ in the former, which is the crux of the problem at hand.  In Eq.\ (\ref{2}) the distribution of a parton with momentum $q$ at the medium surface is related to the parton distribution $f_i(k)$ at the point of creation whose dependence on $k$ is known \cite{sg}.  The relationship between $f_i(k)$ and $F_i(q, \phi, b)$ is complicated in view of Eqs.\ (\ref{2}) and (\ref{3}), but has been simplified to Eq.\ (\ref{6}) by the introduction of $\xi$.  In terms of $\xi$, the degradation of momentum from $k$ to $q$ can be written as a simple exponential for $6<k<12$ GeV/c, i.e.,
\begin{eqnarray}
G(k,q,\xi)=q\delta(q-ke^{-\xi}) ,  \label{42}
\end{eqnarray}
whose justification based on an approximation of QCD result  is given in Ref. \cite{hy2}.  Using this in Eq.\ (\ref{6}) yields the simple connection
\begin{eqnarray}
F_i(q,\xi)=q^2e^{2\xi} f_i(qe^\xi) .  \label{43}
\end{eqnarray}
With this substituted into Eqs.\ (\ref{36}) and (\ref{38}) in place of $F_i(q)$, we obtain the TS and SS contributions to the pion distribution as functions of $p_T$ and $\xi$
\begin{eqnarray}
\rho_1^{\rm TS+SS}(\pt,\xi) = \int{dq\over q}\sum_i F_i(q,\xi) H_i(q,\pt) ,  \label{44}
\end{eqnarray}
where $H_i(q,\pt)$ is as noted after Eq.\ (\ref{38}).  To determine the dependence on $\phi$ and $b$, it is necessary to make one last transform according to Eq.\ (\ref{5}), using what we have learned about $P(\xi, \phi, b)$.  It should now be clear that whereas Eqs.\ (\ref{43}) and (\ref{44}) describe the dynamical processes of hard parton scattering, momentum degradation and hadronization, the geometrical complication is contained in $P(\xi, \phi, b)$.  The validity of this separation remains to be verified by comparison with data in a way that can best reveal the role of $\xi$.  

We calculate the nuclear modification factor
\begin{eqnarray}
R_{AA}^{\pi}(\pt,\phi,c)={dN_{AA}^{\pi}/d\pt d\phi\over N_{\rm coll} dN_{pp}^{\pi}/d\pt} ,  \label{45}
\end{eqnarray}
where $N_{\rm coll}$ is the average number of binary collisions and $dN^{\pi}_{\rm pp}/dp_T$ is the pion distribution in $pp$ collisions.  For AuAu collisions we include the TT component, although it is small compared to the TS and SS components for $p_T >4$ GeV/c.  In Ref.\ \cite{af} good quality data are given for $4<p_T<5$ GeV/c, six bins of $\phi$ and five bins of $c$.  At higher $p_T$ the errors are larger.  We calculate the same quantities for $p_T=4.5$ and 7.5 GeV/c, and for similar ranges of $\phi$ and $c$.  An important step we take is to present our results as functions of $\bar\xi(\phi,c)$.

\begin{figure}[tbph]
\includegraphics[width=0.4\textwidth]{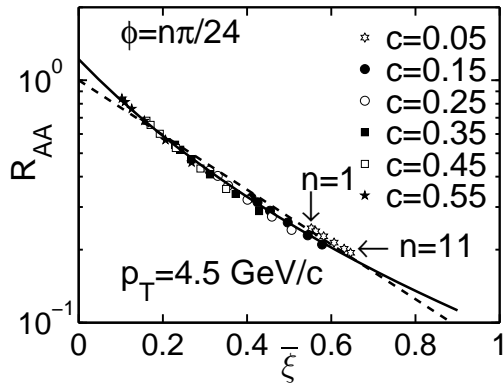}
\caption{Theoretical result showing scaling behavior of $R_{AA}$ at $p_T=4.5$ GeV/c for six values of centrality. The points for different $\phi$ values between 0 and $\pi/2$ have the same symbol when the centralities are the same,  with $\phi$ increasing from left to right with $n=1,3,\cdots, 11$. The solid line is a fit by Eq.\ (\ref{46}) and the dashed line is an exponential fit by $\exp(-2.6\bar\xi)$.}
\end{figure}

Since for each pair of values of $\phi$ and $c$, we can calculate $R^{\pi}_{AA}(p_T, \phi, c)$ and $\bar \xi(\phi, c)$, they can therefore be plotted one versus the other, as in Fig.\ 3.  We have done this for six values of $\phi$ at $n\pi/24$, $n=1, 3, \cdots , 11,$ and six values of $c$ at $0.05m, m= 1, 3, \cdots , 11$, with $p_T$ fixed at 4.5 GeV/c.  Points for various $\phi$ with the same $c$ have the same symbol; they are spread out from left to right with increasing values of $\phi$.  All 36 points lie remarkably well on one universal curve.  It means that different experimental conditions on $\phi$ and $c$ yield the same value of $R^{\pi}_{AA}$ so long as their values of $\bar\xi(\phi, c)$ are the same.  The nature of the behavior of $R^{\pi}_{AA}(\bar\xi)$ depends on the details of the dynamics.  It can be approximated by an exponential \begin{eqnarray}
R_{AA}^\pi(\bar\xi)\left|_{\pt=4.5}\approx \exp (-2.6\bar\xi)\right. ,  \label{45a}
\end{eqnarray}
 as shown by the dashed straight  line in Fig.\ 3, although a better fit is
\begin{eqnarray}
R_{AA}^\pi(\bar\xi)\left|_{\pt=4.5} = 1.22(1+\bar\xi/0.69)^{-2.86} \right. ,  \label{46}
\end{eqnarray}
shown by the solid line.  That is a simple formula for a wide range of $\phi$ and $c$.

\begin{figure}[tbph]
\includegraphics[width=0.4\textwidth]{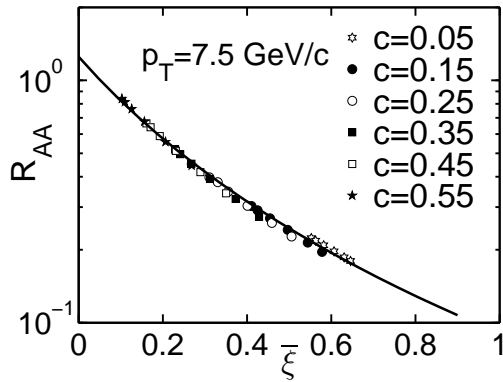}
\caption{Same as for Fig.\ 3 but for $p_T=7.5$ GeV/c. The solid line is a fit by Eq.\ (\ref{47})}
\end{figure}

Similar calculation can be done for $p_T=7.5$ GeV/c.  Fragmentation dominates at such a high $p_T$.  We include all contributions to the hadronization with the result shown in Fig.\ 4.  Again, good scaling is found for all $\phi$ and $c$.  The solid line shows a fit of the $\bar\xi$ dependence by
\begin{eqnarray}
R_{AA}^\pi(\bar\xi)\left|_{\pt=7.5} = 1.25(1+\bar\xi/0.58)^{-2.62} \right. .  \label{47}
\end{eqnarray}
A comparison with Eq.\ (\ref{46}) indicates that there is no significant change in $R^\pi_{AA}(\bar\xi)$ as $p_T$ is increased from 4.5 to 7.5 GeV/c.  Average values of the two sets of parameters can yield acceptable fits of all points calculated.  Thus we learn that there is an approximately exponential suppression of $R_{AA}$ in terms of the path length $\xi$ for all $\phi$ and all centrality $<60\%$, for $p_T>4$ GeV/c, bearing in mind that $\bar\xi$ is no greater than 0.65.  The degree of suppression is no more than a factor of 5, which is a familiar fact.  In Figs. 3 and 4, we see clearly how that suppression is related to one another at different $\phi$ and $c$.  

To compare our calculations with the data we have one basic parameter to vary, which is $\gamma$ in Eq.\ (\ref{4}) that relates $\xi$ to $\ell(x_0, y_0, \phi, b)$, defined in Eq.\ (\ref{15}).  That integral for  $\ell(x_0, y_0, \phi, b)$ is along the path of the hard-parton trajectory weighted by the local density $D(x, y)$, which is proportional to the longitudinal length at $(x, y)$.  With the normalization factor being absorbed by $\gamma$, $\ell$ itself has the characteristic of length, and $\gamma$ behaves as inverse length, though all length are in units of $R_A$.  The dimensionless $\xi=\gamma\ell$ is the dynamical path length, in terms of which the momentum degradation can be expressed as $e^{-\xi}$.  Our formalism does not provide a way to calculate $\gamma$, but it does account for the geometrical details.  In QCD energy loss can be calculated, but without the details of the medium.  If the scaling behavior that we have found can also be found in the data, then the value of $\gamma$ can be extracted, giving justification to our procedure of condensing all the geometrical complications into $P(\xi, \phi, b)$.

\begin{figure}[tbph]
\includegraphics[width=0.4\textwidth]{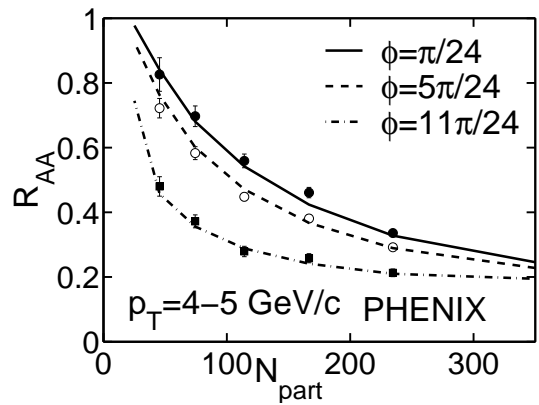}
\caption{$R_{AA}$ vs $N_{\rm part}$ for 3 values of $\phi$ at $\pt$=4-5 GeV/c. The data are from Ref.\ \cite{af}. The lines are from calculation in the recombination model.}
\end{figure}

In Fig.\ 5 we show the data from Ref.\ \cite{af} on $R_{AA}$ vs $N_{\rm part}$ at $4<p_T<5$ GeV/c for 3 values of $\phi$.  We adjust the parameter $\gamma$ to fit the data and obtain
\begin{eqnarray}
\gamma=0.11 .   \label{48}
\end{eqnarray}
There is another parameter that has not been mentioned so far.  It is the lower limit $t_0$ of the integral in Eq.\ (\ref{15}).  It can differ slightly form $0$ to take into account the absence of immediate effect of energy loss at the very beginning of a hard-parton trajectory.  A non-zero $t_0$ shortens the effective path length and can increase $R_{AA}$ at small $N_{\rm part}$ and small $\phi$.  A better fit of the data in that region is achieved by using $t_0=0.025$, which corresponds to less than 0.2 fm, a value small enough so that it does not play a significant role in the general description of the problem.  All figures shown above are results of calculations done with $\gamma$ and $t_0$ given these values.  Another way of presenting the results is to plot $R_{AA}$ versus $\phi$ for different centralities.  That is done in Fig.\ 6, showing good agreement with data \cite{af}.

\begin{figure}[tbph]
\includegraphics[width=0.4\textwidth]{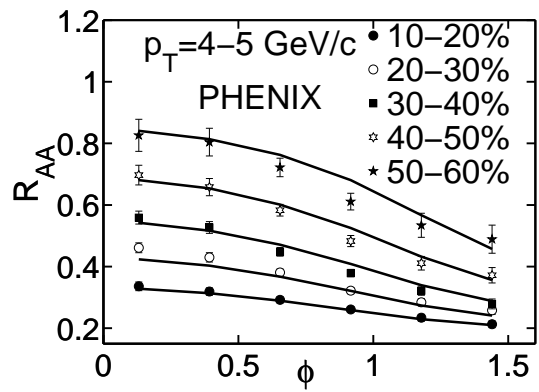}
\caption{$R_{AA}$ vs $\phi$ for 5 bins of centrality at $\pt$=4-5 GeV/c. The data are from Ref.\ \cite{af}. The lines are from calculation in the recombination model.}
\end{figure}

As we can see from these two figures, the data plotted in terms of $\phi$ and $c$ vary over wide ranges of values, exhibiting no organized simplicity.  It is therefore important to replot $R_{AA}$ versus $\bar\xi(\phi, c)$ to check the scaling behavior found in our theoretical study.  Fig.\ 7 shows the data replotted in that format; the solid line is a direct transfer from the solid line in Fig.\ 3 that fits the theoretical points by Eq.\ (\ref{46}).  It is a way to show the relationship between the two figures, but in actuality the theoretical and experimental points agree with one another better than how the analytic formula can represent them.  Given the experimental errors, the scaling behavior is evidently in the data.

\begin{figure}[tbph]
\includegraphics[width=0.4\textwidth]{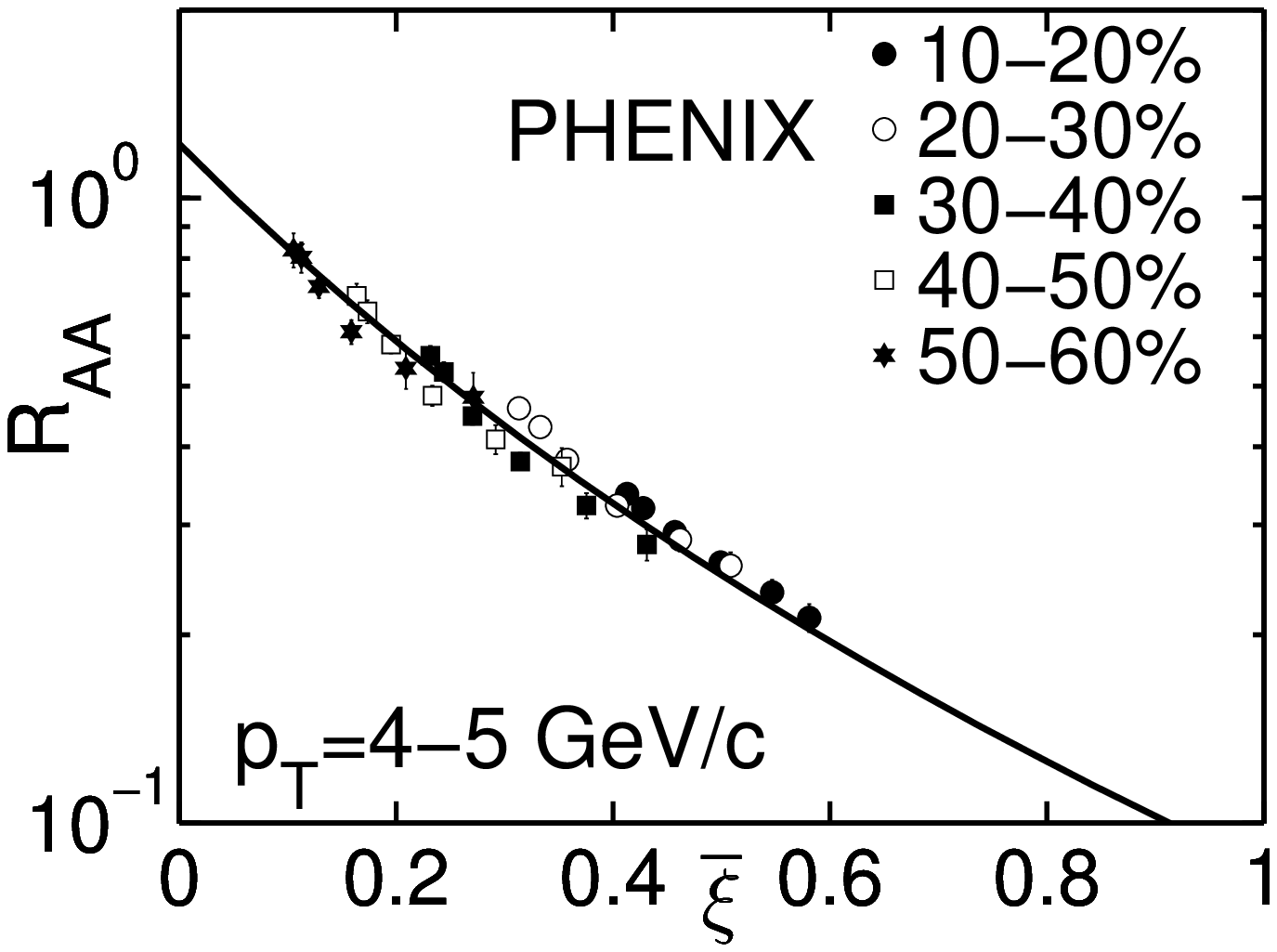}
\caption{Experimental data on $R_{AA}$ for $\pt$=4-5 GeV/c from Ref.\ \cite{af} are plotted against the scaling variable $\bar\xi$. The $\phi$ values are the same as described in Fig.\ 3. The solid line is the same as the solid line in Fig.\ 3 that represents the  result from theoretical calculation.}
\includegraphics[width=0.4\textwidth]{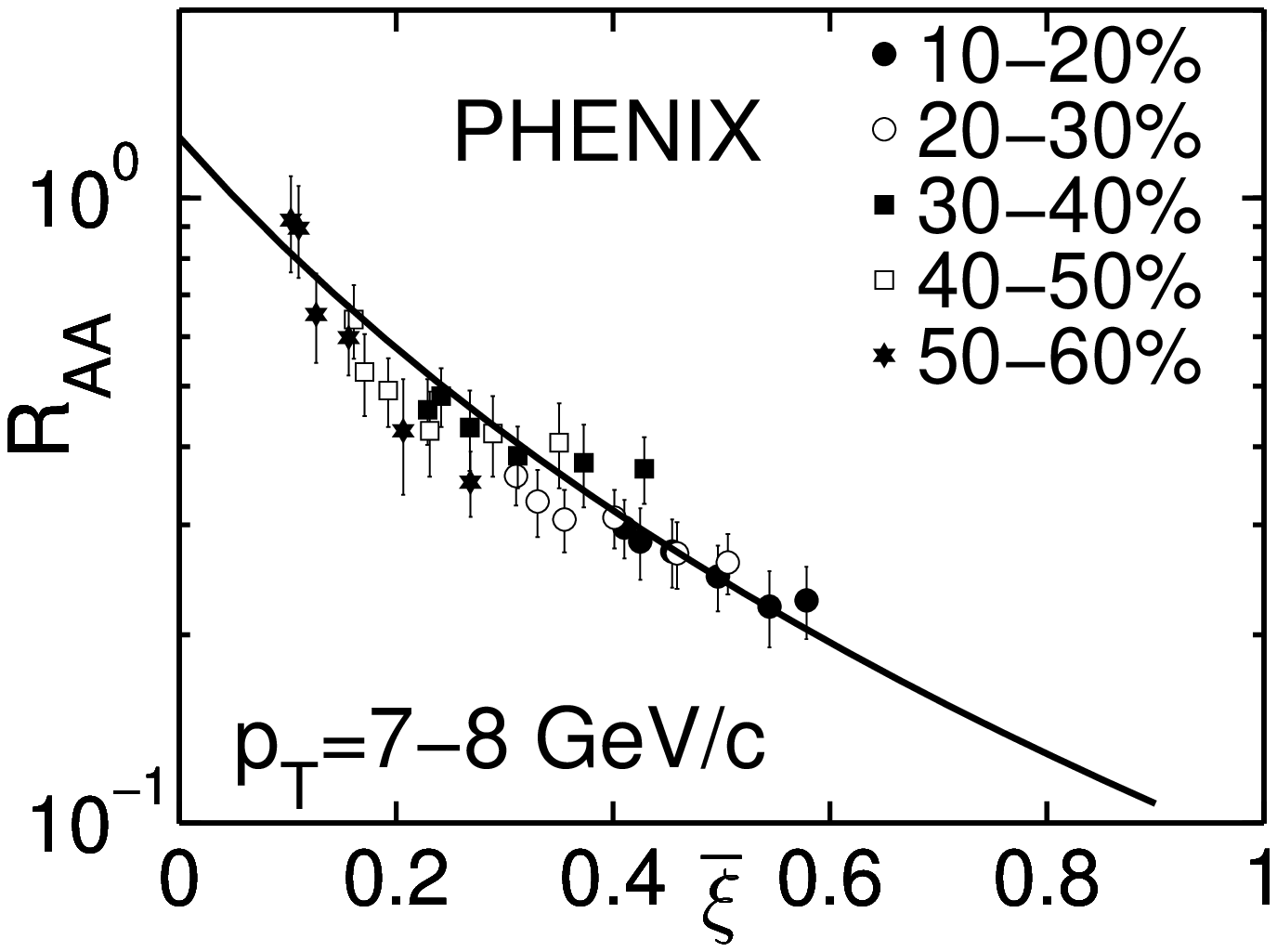}
\caption{Experimental data on $R_{AA}$ for $\pt$=7-8 GeV/c from Ref.\ \cite{af} are plotted against the scaling variable $\bar\xi$. The $\phi$ values are the same as described in Fig.\ 3. The solid line is the same as the solid line in Fig.\ 4 from theoretical calculation.}
\end{figure}

For $p_T$ in the 7-8 GeV/c range we show the data only in the scaling format, as in Fig.\ 8.  The solid line is a reproduction of the solid line in Fig.\ 4, described by Eq.\ (\ref{47}).  Within errors which are large, the agreement between theory and experiment is good in both the magnitude and the $\bar\xi$ dependence.  No parameters have been adjusted.  Thus we learn not only that scaling is robust, but also that the dependence on $\bar\xi$ changes with $p_T$ mainly because the SS component of hadronization becomes most dominant at higher $p_T$, since no other factor besides $H_i(q, p_T)$ in Eq.\ (\ref{44}) depends on $p_T$.  The medium effect is primarily contained in $F_i(q, \xi)$.

It should be mentioned that the path length dependence has also been investigated in Ref.\ \cite{af}. Scaling behavior has been found in different variables for different $\pt$ ranges. Their definition of $\rho L_{xy}/\rho_{\rm cent}$ in fm is closest in spirit to our dimensionless $\bar\xi$. The linear dependence of $\ln R_{AA}$ on the path length $L$ and $p_T^{-1/2}$ has also been checked experimentally in Refs.\ \cite{la,lw}, as predicted in QCD \cite{dk}. Since our $\bar\xi(\phi,c)$ is far more detailed than $L$ and our $\pt$ range includes the contribution from thermal-shower recombination, our result cannot be related simply to any QCD formula.

It is interesting to show how the $\bar\xi$ dependence can be obtained in a more direct and transparent way, when an approximation is made at high $p_T$.  The hard-parton distribution $f_i(k)$ has been parametrized in Ref. \cite{sg} in the form
\begin{eqnarray}
f_i(k)\propto (1+k/k_i)^{-\beta_i}   \label{49}
\end{eqnarray}
where for $i=g\ (u), k_i=1.77\ (1.46)$ GeV/c and $\beta_i=8.61\ (7.68)$. For simplicity, we approximate the formula by a simple power law for $k>7$ GeV/c
\begin{eqnarray}
f_i(k)\propto k^{-\beta'_i} ,   \label{50}
\end{eqnarray}
where $\beta'_g=7.8$ and $\beta'_u=7.08.$ The advantage is that $F_i(q,\xi)$ becomes, from Eq.\ (\ref{43}),
\begin{eqnarray}
F_i(q,\xi) \propto (qe^\xi)^{-\tau_i} , \qquad \tau_i=\beta'_i-2 , \label{51}
\end{eqnarray}
which renders the $q$ and $\xi$ dependencies separable.  We can then perform the $q$ integration in Eq.\ (\ref{44}) independent of $\xi$, and obtain the gluon and u-quark contributions separately as
\begin{eqnarray}
R_{AA}^i(\pt,\phi,c) \propto \int d\xi P(\xi,\phi,c)e^{-\xi\tau_i} . \label{52}
\end{eqnarray}
The scaling property of $P(\xi, \phi, c)$, as described by Eq.\ (\ref{25}), can now be used to yield
\begin{eqnarray}
R_{AA}^i(\pt,\bar\xi) \propto \int dz \psi(z)e^{-z\xb\tau_i} . \label{53}
\end{eqnarray}
which can simply be evaluated in conjunction with Eq.\ (\ref{27}).  The result is shown in Fig.\ 9 for gluon in (green) solid line and for $u$ quark in (red) dashed line, the normalizations for which are freely adjusted.  It should be noted that the curve in Fig.\ 7 is from complete calculation taking the contributions from all partons into account without using explicitly the scaling property of $P(\xi, \phi, c)$, while the two curves in Fig.\ 9 are obtained through the approximation of Eq.\ (\ref{50}) so as to exhibit directly the roles of (a) the parton distribution function $f_i(k)$, (b) the exponential relationship between $k$ and $q$, and (c) the scaling function $\psi(z)$.  All three of these properties are the main content of the integrand in Eq.\ (\ref{53}) that shows clearly the three basic aspects of the physics that give rise to the observed behavior of $R_{AA}$.

\begin{figure}[tbph]
\includegraphics[width=0.4\textwidth]{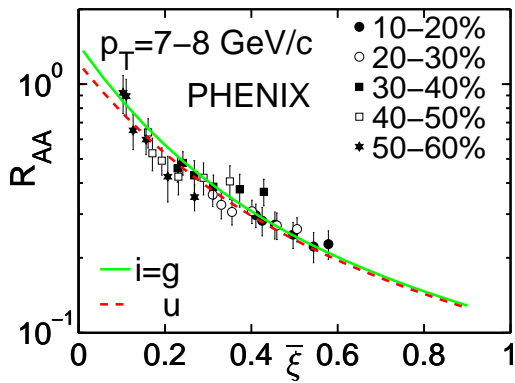}
\caption{(Color online) The same data as shown in Fig.\ 8 for $\pt$=7-8 GeV/c. The (green) solid line is for gluon fragmentation only and the (red) dashed line is for $u$-quark fragmentation; the normalizations for both contributions are arbitrarily adjusted so as to show the $\bar\xi$ dependence compared to data.}
\end{figure}

It is of interest to point out the implication of the two factors in the integrand of Eq.\ (\ref{52}), or (\ref{53}).  The factor $\exp(-\xi\tau_i)$ is basically an expression of the hard-parton distribution $f_i(k)$ in terms of $\xi$, indicating that the probability of a hard parton contributing at large $\xi$ to an observed pion is exponentially suppressed.  $P(\xi, \phi, c)$ gives the probability of creating a hard parton at $\xi$ from the surface.  Their product, expressed as $\psi(z)\exp(-z\bar\xi\tau_i)$ in Eq.\ (\ref{53}), shows that, although the mean $z$ is 1, the small $z$ portion dominates due to higher suppression at large $z$.  It means that most of the partons that contribute to pion production are created near the surface.  If the result of the integration over $z$ can be approximated by $\exp(-\hat z\bar\xi\tau_i)$ with $\hat z=0.5$, we see that the dependence of $R_{AA}^\pi(\bar\xi)$ on $\bar\xi$ is roughly the exponential behavior given by Eq.\ (\ref{45a}).

\section{BACK-TO-BACK DIJETS}

Having seen how the dynamical path length $\xi$ plays a crucial role in determining the general behavior of $R_{AA}$, it is natural to extend the study to consider back-to-back dijets, whose properties clearly depend on the distances that the two generating hard partons traverse in the medium.  That problem has been studied previously in Ref.\ \cite{hy2} without taking into account azimuthal dependence explicitly.  We now can investigate the effect of $\xi$ scaling and find a more transparent way to show the dependence of away-side yield per trigger on $\phi$ and $c$.

Using the same notation as in [4] where $p_t$ and $p_b$ denote the momenta of the trigger and the associated particle on the away side, respectively, at mid-rapidity, we have by simple generalization  from Eq.\ (\ref{1})
\begin{eqnarray}
&&{dN_{AA}\over p_tp_bdp_tdp_bd\phi}(b)=\rho_2(p_t,p_b,\phi,b) \nonumber \\
&&=\int {dq_1\over q_1}{dq_2\over q_2}\sum_i F_i(q_1,q_2,\phi,b)H_i(q_1,q_2,\phi,b) ,  \label{54}
\end{eqnarray}
where $q_1$ is the parton momentum on the near-side surface, and $q_2$ on the away-side surface.  They are related to the momentum $k$ at the creation point by $G(k, q_j, \xi_j), \ j=1, 2.$  Eq.\ (\ref{4}) is now generalized to 
\begin{eqnarray}
&&P(\xi_1,\xi_2,\pb = \int dx_0dy_0 Q(x_0,y_0,b)\qquad\nonumber \\
&&\times\delta(\xi_1-\gamma\ell_1(x_0,y_0,\phi,b))\delta(\xi_2-\gamma\ell_2(x_0,y_0,\phi,b)) , \label{55}
\end{eqnarray}
where $\ell_1(x_0, y_0, \phi, b)$ is as defined in Eq.\ (\ref{15}), and $\ell_2(x_0, y_0, \phi, b)$ is defined similarly but in the opposite direction with the upper limit of integration replaced by
\begin{eqnarray}
t_2(x_0, y_0, \phi, b)=t_1(x_0, y_0, \phi+\pi, b) .  \label{56}
\end{eqnarray}
In our calculation below we shall set the lower limit of both integrals at $t=0$, ignoring the small $t_0$ mentioned after Eq.\ (\ref{48}).  As in Eqs.\ (\ref{5}) and (\ref{6}) we have
\begin{eqnarray}
F_i(q_1,q_2,\phi,b)=\int d\xi_1d\xi_2 P(\xi_1,\xi_2,\pb F_i(q_1,q_2,\xi_1,\xi_2) ,  \label{57} \\
F_i(q_1,q_2,\xi_1,\xi_2) = \int dk k f_i(k) G(k,q_1,\xi_1) G(k,q_2,\xi_2) . \label{58}
\end{eqnarray}
Computation of $\rho_2(p_t, p_b, \phi, b)$ can now be performed from the equations given above without questioning the scaling properties.  However, we have learned from the preceding section that considerable clarity can be gained by presenting the result in terms of the scaling variable $\bar\xi$.

Our next step in that direction is therefore to study the properties of $P(\xi_1, \xi_2, \phi, b)$.  We define
\begin{eqnarray}
z_j=\xi_j/\bar\xi(\pb , \quad j=1,2,  \label{59}
\end{eqnarray}
where $\bar\xi(\phi, b)$ is the same as determined in Sec.\ 3 and shown in Fig.\ 1.  It should be noted that although Eq.\ (\ref{23}) expresses $\bar\xi(\phi,b)$ in terms of the near-side path length $\ell_1(x_0, y_0, \phi, b)$, it is the same as for the away side, since $(x_0, y_0)$ is integrated over the entire overlap so replacing $\phi$ by $\phi+\pi$ does not lead to any difference.  We now ask whether $P(\xi_1, \xi_2, \phi, b)$ has the scaling behavior 
\begin{eqnarray}
P(\xi_1, \xi_2, \phi, b)=\Psi(z_1,z_2)/\bar\xi^2(\pb , \label{60}
\end{eqnarray}
where
\begin{eqnarray}
\int dz_{2,1} \Psi(z_1,z_2)=\psi(z_{1,2}) .   \label{61}
\end{eqnarray}
For $\phi=\pi/24$ and $c=0.05$, we have $\bar\xi=0.55$ and we can plot $P(\xi_1, \xi_2)$ in the format of $\Psi(z_1, z_2)$ vs $(z_1, z_2)$, as shown in Fig.\ 10.  It is essentially symmetric in $z_1$ and $z_2$, as it should, and peaks near the boundary of a maximum $z_1+z_2$.  The existence of such a boundary is due to the finite size of the medium that puts an upper limit on $\xi_1+\xi_2$.  Fig.\ 11 shows the maximum $z_2$ as a function of $z_1$.  Evidently, the symmetry in $z_1$ and $z_2$ suggests the use of the variables 
\begin{eqnarray}
z_\pm = z_1 \pm z_2 ,  \label{62}
\end{eqnarray}
so that $z_+^{\rm max}=2.4$.  We show in Fig.\ 12 the projections of $\Psi(z_+, z_-)$ on $z_+$ for 4 values of $z_-$.  The four curves are in essence very similar although in detail they cannot be identical because $z_1$ and $z_2$ must both be positive so $z_+$ has different lower limits depending on $z_-$.

\begin{figure}[tbph]
\includegraphics[width=0.4\textwidth]{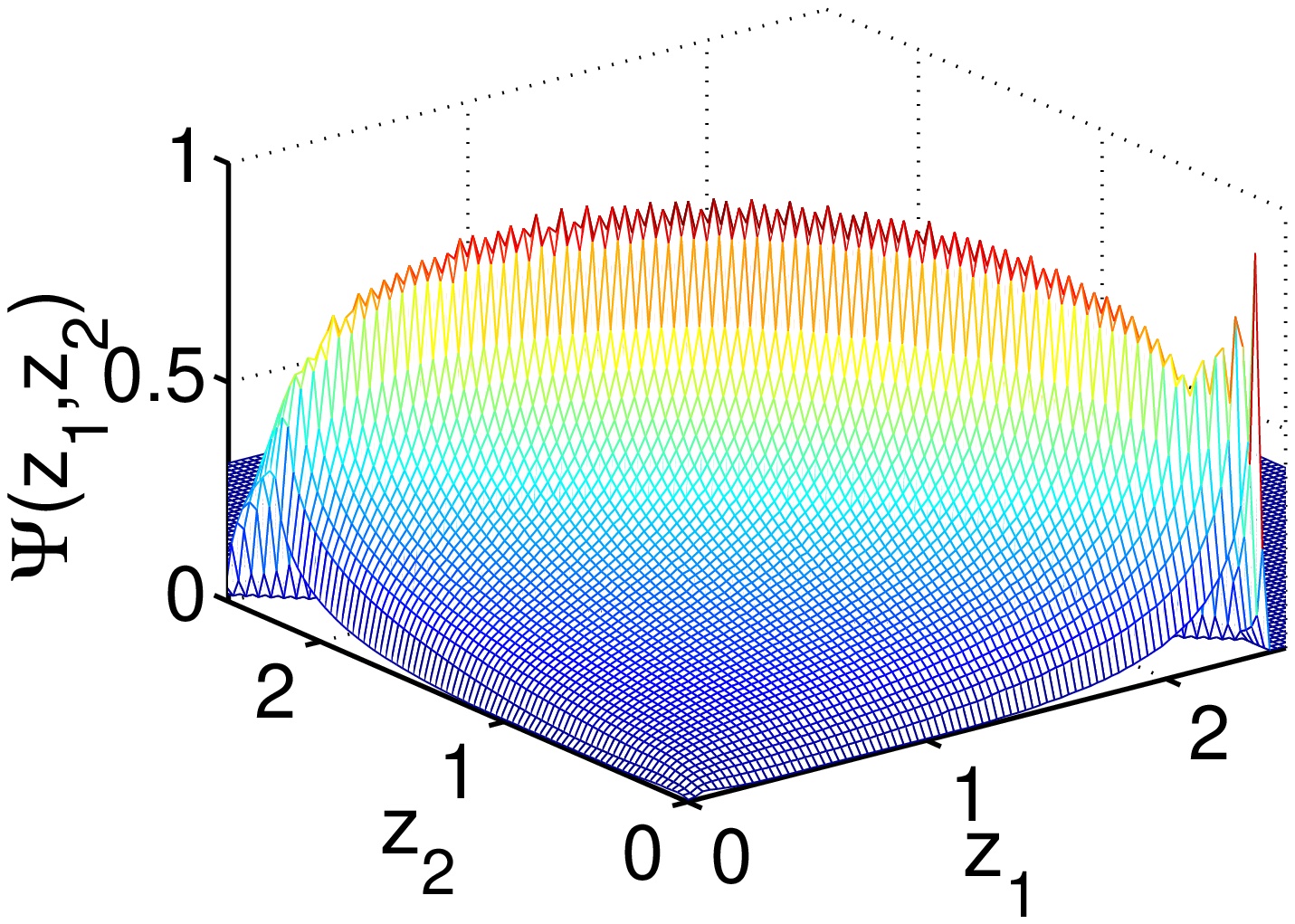}
\caption{(Color online) A 3D plot of the scaling function $\Psi(z_1,z_2)$ for $c=0.05$ and $\phi=\pi/24$.}
\includegraphics[width=0.35\textwidth]{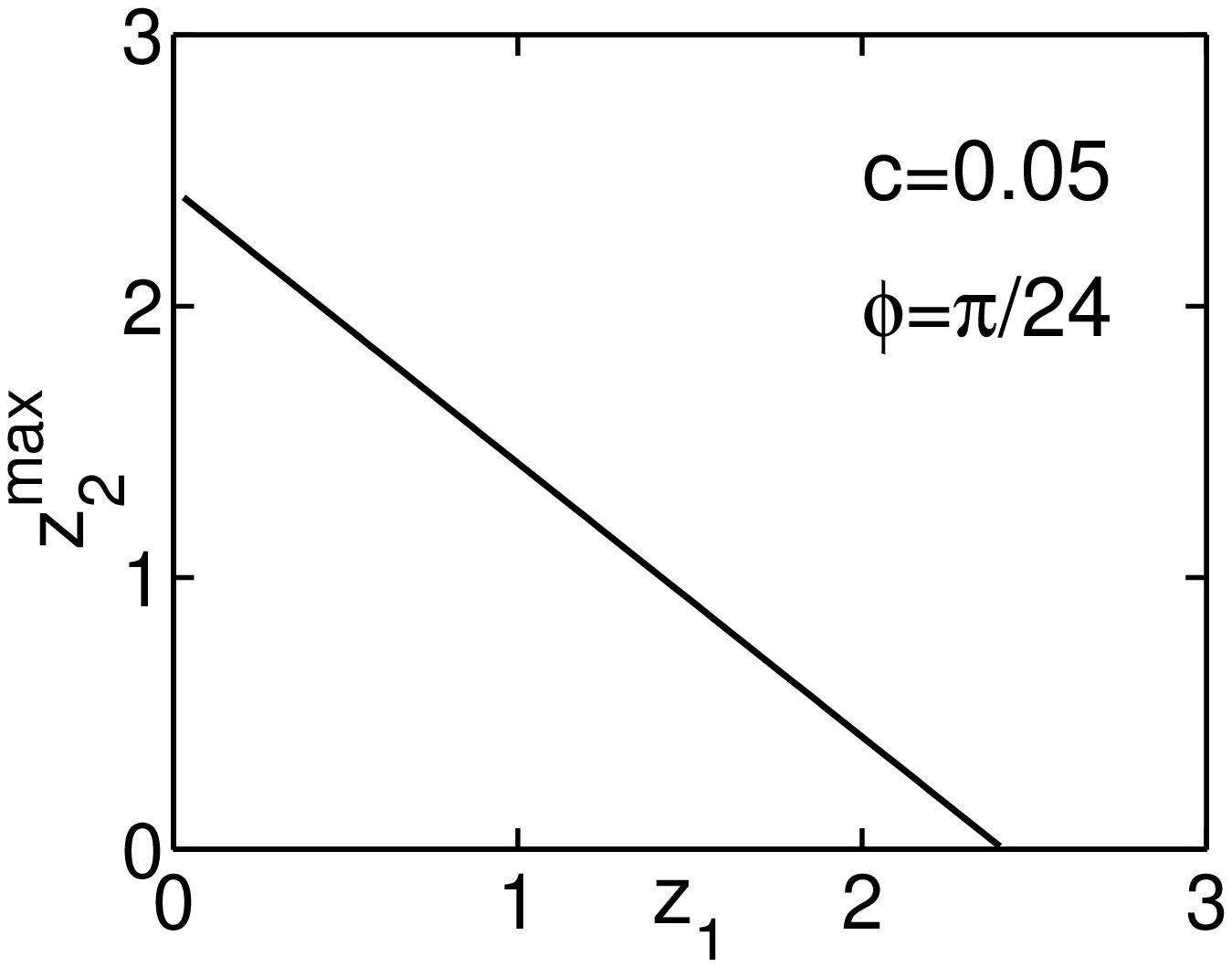}
\caption{Maximum $z_2$ vs $z_1$ for $c=0.05$ and $\phi=\pi/24$.}
\end{figure}

\begin{figure}[tbph]
\includegraphics[width=0.4\textwidth]{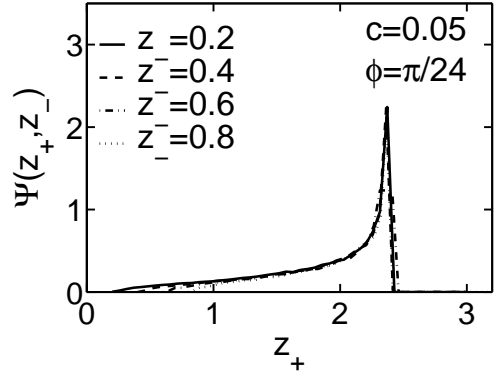}
\caption{The scaling function $\Psi(z_+,z_-)$ vs $z_+$ for four values of $z_-$ at $c=0.05$ and $\phi=\pi/24$.}
\end{figure}

Figures 10-12 are for $\phi=\pi/24$ and $c=0.05$.  To show scaling it is necessary to do the calculation for other values of $\phi$ and $c$ and demonstrate universality.  We have done that and found plots like Fig.\ 12 that have almost no perceptible differences among them and are thus not exhibited.  We conclude that $P(\xi_1, \xi_2, \phi, b)$ does satisfy scaling as described by Eq.\ (\ref{60}).

We can now return to the dijet spectra and present the results on yield per trigger defined as
\begin{eqnarray}
Y_{\pi\pi}^{\rm away}(p_t,p_b,\bar\xi)={\rho_2(p_t,p_b,\pb \over \rho_1(p_t, \pb}  ,  \label{63}
\end{eqnarray}
which is plotted as a function of $\bar\xi$ in Fig.\ 13 for $p_t=8$ GeV/c and $p_b=4, 6, 8$ GeV/c.  The symbols are the same as those in Fig.\ 3, representing $\phi=n\pi/24$ and $c=0.05m$.  There is clearly good scaling behavior in $\bar\xi$.  They are the results of direct calculation of $\rho_2(p_t, p_b, \phi, c)$ without using the properties of $\Psi(z_1, z_2)$ in Figs.\ 10 and 12.  The dependence on $\bar\xi$ is similar among the three sets of $p_b$ values, although for $p_b=4$ GeV/c the decrease with $\bar\xi$ is slightly slower than for 6 and 8 GeV/c.  The normalization of each set is due to many factors, among which are obviously the fragmentation functions $D_i(p_b/q_2)$ on the away side contained in $H_i(q_1, q_2, p_t, p_b)$ in Eq.\ (\ref{54}); they can directly account for the decrease with increasing $p_b$ if $q_2$ were held fixed.  Since $\xi_2$ affects the dependence on $q_2$, it is understandable why the $\bar\xi$ dependence of $Y_{\pi\pi}^{\rm away}(p_t, p_b, \bar\xi)$ at $p_b=4$ GeV/c can differ from that at $p_b = 8$ GeV/c, though only slightly.

\begin{figure}[tbph]
\includegraphics[width=0.4\textwidth]{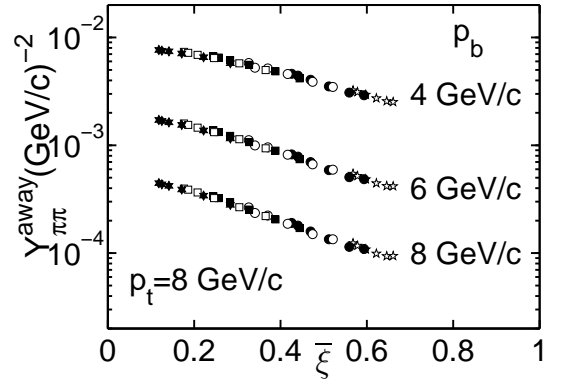}
\caption{Away-side yield per trigger at trigger momentum $p_t=8$ GeV/c and for three values of the associated-particle momentum $p_b$. The symbols are the same as in Fig.\ 3 for various values of $c$ and $\phi$.}
\end{figure}

Another way of presenting the results shown in Fig.\ 13 is to exhibit the $p_b$ dependence for fixed values of $\bar\xi$. In order to compare with data we relate $Y_{\pi\pi}^{\rm away}(p_t,p_b)$ to the trigger-normalized fragmentation function  $D_{\pi\pi}(z_T)$ \cite{whs}, where $z_T=p_b/p_t$, by 
 \begin{eqnarray}
D_{\pi\pi}(z_T,\bar\xi)=p_tp_bY_{\pi\pi}^{\rm away}(p_t,p_b,\bar\xi),  \label{64a}
\end{eqnarray}
so that the integral of both sides over $z_T$ gives the same total yield of pions on the away-side of a fixed trigger pion at $p_t,\phi,c$. In Fig.\ 14 we show the result for three values of $\bar\xi$. The $z_T$ dependence does not change very much in the range $0.5<z_T<1.0$ corresponding to $4<p_b<p_t=8$ GeV/c, when $\bar\xi$ is increased from 0.1 to 0.5. There are recent data from STAR on $\pi^0$-$h^{\pm}$ correlation on the away side for $8<p_t<16$ GeV/c and $0.35<z_T<0.95$ at 0-10\% centrality \cite{bia}. They are shown in Fig.\ 14 for qualitative comparison with our calculated result, although the large variation in trigger momentum may render quantitative comparison meaningless. The general $z_T$ behaviors in theory and experiment are consistent.

\begin{figure}[tbph]
\includegraphics[width=0.4\textwidth]{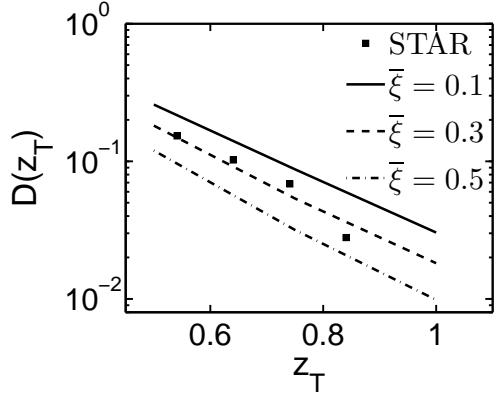}
\caption{Trigger-normalized fragmentation function on the away side. Data are from Ref.\ \cite{bia}}
\end{figure}

The value of $\bar\xi(\phi, c)$ is determined from geometrical considerations, apart from the multiplicative factor $\gamma$.  The dependence of the yield on $\bar\xi$ is, however, rooted in the dynamics of hard scattering and energy loss.  To shed some light on the role played by scaling, we go first to Eq.\ (\ref{58}) and with the help of Eq.\ (\ref{42}) write it in the form 
\begin{eqnarray}
F_i(q_1,q_2,\xi_1,\xi_2)&=&\left[ k^3f_i(k)\right]_{k=(q_1q_2e^{\xi_1+\xi_2})^{1/2}} \nonumber \\
&&\times\delta(q_1e^{\xi_1}-q_2e^{\xi_2}) .  \label{64}
\end{eqnarray}
Using Eqs.\ (\ref{57}) and (\ref{60}) we obtain
\begin{eqnarray}
&&F_i(q_1,q_2,\bar\xi)=\int dz_1dz_2\Psi(z_1,z_2) \nonumber \\
&&\times\left[ k^3f_i(k)\right]_{k=(q_1q_2e^{\bar\xi(z_1+z_2)})^{1/2}}\delta(q_1e^{\bar\xi z_1}-q_2e^{\bar\xi z_2}). \quad  \label{65}
\end{eqnarray}
For symmetric dijets, $p_t=p_b=8$ GeV/c, $q_1$ and $q_2$ are on average equal, so $z_1\approx z_2$.  Eq.\ (\ref{65}) can then be approximated by
\begin{eqnarray}
F_i(q_1,q_2,\bar\xi)\propto\int dz_+dz_-\Psi(z_+,z_-) e^{(3-\beta'_i)\bar\xi z_+/2}\bar\xi^{-1}\delta(z_-) .  \label{66}
\end{eqnarray}
where Eq.\ (\ref{50}) has been used, and multiplicative factors involving $q_1$ and $q_2$ are omitted from the expression.  Writing $\Psi(z_+, z_-=0)$ as $\Psi(z_+)$, we then have
\begin{eqnarray}
\rho_2(p_t,p_b, \bar\xi)_{p_t=p_b} \propto {1\over \bar\xi} \int dz_+ \Psi(z_+) e^{(3-\beta'_i)\bar\xi z_+/2} .  \label{67}
\end{eqnarray}
where $\Psi(z_+)$ is very nearly the solid line in Fig.\ 12, rendering the $\bar\xi$ dependence explicit.  Since $\Psi(z_+)$ is very different from $\psi(z), \rho_2(p_t, p_b, \bar\xi)$ is significantly different from $\rho_1(p_T, \bar\xi)$, which is of the form given in Eq.\ (\ref{53}).  In both cases there is scaling in $\bar\xi$ by virtue of the scaling properties of $P(\xi, \phi, b)$ and $P(\xi_1, \xi_2, \phi, b)$.

\section{TWO-JET RECOMBINATION AT LHC}

In heavy-ion collisions at CERN-LHC the density of hard partons created is so high that the shower partons from neighboring jets can recombine to form hadrons.  It is found in Ref. \cite{hy4} that large $p/\pi$ ratio, as high as  20, can be a signature of such hadronization processes.  Here we examine a different signal that may show up even more dramatically in the preliminary data collected in the initial phase of the operation of LHC.  In simple terms the $\bar\xi$ scaling of $R_{AA}$ may be badly broken, and $R_{AA}$ itself can be very large.

We consider pion production only at high $p_T$ where thermal partons can be ignored, but not so high that the hard-parton density becomes low.  For definiteness, we set $p_T=10$ GeV/c for $\sqrt{s}=5.5$ TeV in PbPb collisions.  As before, we consider only hadron production at midrapidity.  The 2-jet contribution to the inclusive distribution is a simple generalization of Eq. (\ref{1})
\begin{eqnarray}
&&{dN_{AA}^{2j}\over p_Tdp_Td\phi}(b)=\rho_1^{2j}(p_T,\pb  \qquad \nonumber \\
&=&\int{dq\over q}{dq'\over q'}\sum_{i,i'}F_i(q,\pb F_{i'}(q',\pb H_{ii'}(q,q',p_T) , \label{68}
\end{eqnarray}
where $q$ and $q'$ are the momenta of two hard partons at the surface of the medium, both directed at $\phi$, and $H_{ii'}(q, q', p_T)$ describes how those two hard partons hadronize through their separate shower partons that recombine, i.e.,
\begin{eqnarray}
H_{ii'}(q,q',p_T)&=&{1\over p_T^2} \int {dq_1\over q_1}{dq_2\over q_2} S_i^j\left({q_1\over q}\right) S_{i'}^{j'}\left({q_2\over q'}\right) \nonumber \\
&&\times R_\pi^\Gamma(q_1,q_2,p_T). \   \label{69}
\end{eqnarray}
The parton types $i$ and $i'$ are summed over $g,u,d,\bar u$ and $\bar d$, so there are 25 combinations of hard partons, each fragmenting into $j$ and $j'$ shower partons.
The recombination function $R_\pi^\Gamma(q_1,q_2,p_T)$ must now take into account the probability that the two shower partons of types $j$ and $j'$  from two jets can coalesce to form a pion.  Since the two hard partons are parallel, the primary factor is how far apart they are.  Secondarily, the shower partons in a jet may form a cone, so the RF is proportional to the overlap of two such neighboring cones.  There are complications that we do not know well enough to describe quantitatively.  Fortunately, the main effect that we are searching for is independent of the details.  We approximate the RF by
\begin{eqnarray}
R_\pi^\Gamma(q_1,q_2,p_T)=\Gamma R_\pi(q_1,q_2,p_T) , \label{70}
\end{eqnarray}
where $R_\pi(q_1,q_2,p_T)$ is given in Eq. (\ref{32}), and $\Gamma$ is a numerical factor that we shall allow to have widely different values.  More specifically,  we let
\begin{eqnarray}
\Gamma=10^{-m} ,  \qquad m=1, 2, 3.   \label{71}
\end{eqnarray}
If a crude estimate of $\Gamma$ is made by considering the distance between two parallel trajectories, then it would be the ratio of the pion radius to the effective nuclear radius, $R_\pi/R_A^{\rm eff}$.  Taking $R_\pi$ to be 0.6 fm and the $R_A^{\rm eff}\approx 6$ fm (slightly less than $R_A$ due to the predominance of the hard scattering point near the center where $Q(x_0,y_0,b)$ is higher), we get $\Gamma\approx 0.1$.

Transforming the $(\phi, b)$ variables into $\xi$ as in Eq. (\ref{5}), we have 
\begin{eqnarray}
\rho_1^{2j}(p_T,\pb =\int d\xi d\xi' P(\xi,\pb P(\xi',\pb \rho_1^{2j}(p_T,\xi,\xi') , \label{72}
\end{eqnarray}
where
\begin{eqnarray}
\rho_1^{2j}(p_T,\xi,\xi') =\int{dq\over q}{dq'\over q'}\sum_{i,i'}F_i(q,\xi) F_{i'}(q',\xi') H_{ii'}(q,q',p_T) . \label{73}
\end{eqnarray}
The two variables $\xi$ and $\xi'$ are independent, unlike $\xi_1$ and $\xi_2$ in $P(\xi_1, \xi_2, \phi, b)$ considered in the preceding section.  But for any $\phi$ and $b$ there is only one $\bar\xi(\phi, b)$.  Thus KNO scaling implies
\begin{eqnarray}
\rho_1^{2j}(p_T,\pb =\int dz dz' \psi(z) \psi(z') \rho_1^{2j}(p_T,z,z',\bar\xi) , \label{74}
\end{eqnarray}
In our calculation we include also the 1-jet contribution in Eq.\ (\ref{44}), in addition to the 2-jet contribution above.  A number of parameters are changed in going from RHIC to LHC, such as $\sigma^{pp}_{\rm inel}=60\, {\rm mb}, A=207, \omega=14$, and those in $f_i(k)$ \cite{sg}, but we have fixed $\gamma$ at the value given in Eq.\ (\ref{48}) as a reasonable working hypothesis.  
To calculate $R_{AA}^\pi(p_T,\bar\xi)$ we use $dN_{pp}^\pi/p_Tdp_T=8.7\times 10^{-7}$(GeV/c)$^{-2}$ for $p_T=10$ GeV/c at $\sqrt s=5.5$ TeV, getting the 
result  shown in Fig. 15 for $\Gamma=10^{-3}$.  The symbols are the same as those in Figs.\ 3 and 4.  It is clear that there is excellent scaling as before.  However,  the reason for the scaling behavior is the dominance of 1-jet contribution.

\begin{figure}[tbph]
\includegraphics[width=0.4\textwidth]{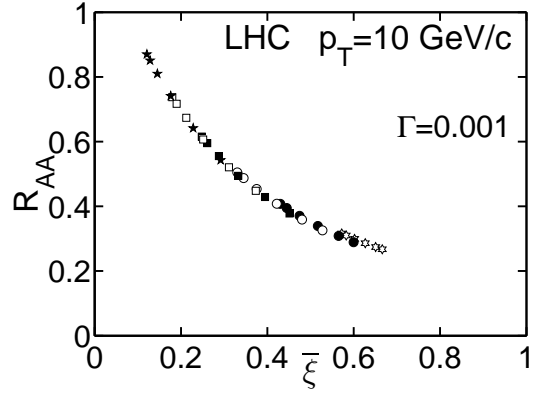}
\caption{Scaling behavior  of $R_{AA}$ at LHC for $\pt$=10 GeV/c and 2-jet recombination overlap factor $\Gamma=0.001$. The symbols are the same as in Fig.\ 3 for various values of $c$ and $\phi$.}
\end{figure}

To show the effect of the 2-jet contribution we repeated the calculation for $\Gamma=0.1$.  The result is shown in Fig.\ 16.  Evidently, scaling is badly broken.  The root cause can be traced to the fact that two jets involve two separate creation points of hard partons, the probability for each being proportional to $N_{\rm coll}$.  Thus the $N_{\rm coll}^2$ dependence of $\rho_1^{2j}(p_T, \xi, \xi')$ is not scaled out by the definition of $R_{AA}^\pi$ in Eq.\ (\ref{45}).  The admixture of $\rho_1^{1j}$ and $\rho_1^{2j}$ therefore depends implicitly on $N_{\rm coll}$ that ruins the $\bar\xi$ scaling.  To confirm our rationale, we have left out $\rho_1^{1j}$ and studied $\rho_1^{2j}/N_{\rm coll}^2$ from Eq.\ (\ref{74}); indeed, we found scaling.

\begin{figure}[tbph]
\includegraphics[width=0.4\textwidth]{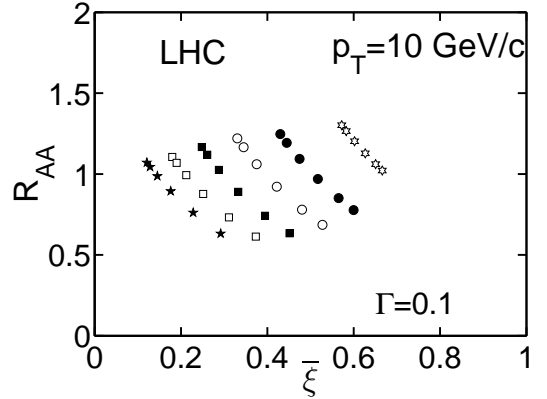}
\caption{Non-scaling behavior  of $R_{AA}$ at LHC for $\pt$=10 GeV/c and  $\Gamma=0.1$. The symbols are the same as in Fig.\ 3 for various values of $c$ and $\phi$.}
\end{figure}

The contrast between Figs.\ 15 and 16 is striking, the only difference in the input being the values of $\Gamma$ used.  Experimental data should readily be able to distinguish the two scenarios, and give hints on the value of $\Gamma$ that we know too little to calculate.  At $p_T$ much higher than 10 GeV/c parton momenta $k$ and $k'$ are higher, so the parton density would be lower and the probability of 2-jet recombination would be suppressed.  At lower $p_T$ the ratio of 2j-to-1j contributions in Pb-Pb collisions without the $\Gamma$ factor increases by a factor of around 30\% when $p_T$ is lowered from 10 to 6 GeV/c.  Hence, if $\Gamma=10^{-3}$, $\bar\xi$ scaling would persist, but if $\Gamma=10^{-1}$, then the scaling violation would also continue to prevail.

Another aspect about Fig.\ 16 that is perhaps more spectacular than the scaling violation is the magnitude of $R_{AA}$ that can exceed 1 for small $\phi$ (left side of each type of symbols). For central collisions ($c=0.05$, open stars) for which $\bar\xi$ is between 0.5 and 0.7, the magnitude of $R_{AA}^\pi$ with dominance by 2-jet recombination for $\Gamma=0.1$ can be four times larger than the 1-jet contribution for $\Gamma=10^{-3}$, both being for pion at $p_T=10$ GeV/c.  The increase is not so large for mid-central collisions at $c=0.55$. Clearly, the $N_{\rm coll}^2$ dependence of $\rho_1^{2j}$ boosts $R_{AA}^\pi$ in central collisions to completely overturn the effect of nuclear suppression.
 Thus the detection of large $R_{AA}^\pi$ (and even larger $R^p_{AA}$ according to Ref.\ \cite{hy4}) at LHC is an immediate signal of two-jet recombination.

\section{CONCLUSION}

For hadron $\pt$ large enough so that the dependence of $R_{AA}$ on centrality $c$ and azimuthal angle $\phi$ is closely related to the propagation of hard partons in a dense medium, our treatment of the geometrical aspect of the problem has led to the finding of a scaling property that not only simplifies considerably the presentation of such dependence, but also offers a clearer picture from which one can learn about the nature of medium effect on jets. In terms of the dynamical path length $\bar\xi(\phi,c)$ there is in $R_{AA}$ an approximate exponential behavior for all $\phi$ and $c$. Based on the results of our calculation in the recombination model, we present the data of Ref.\ \cite{af} in a similar format and found the scaling behavior for $\pt$ at 4-5 and 7-8 GeV/c over wide ranges of $\phi$ and $c$. Furthermore, the dependence on $\bar\xi(\phi,c)$ is as we have obtained theoretically.

Experimental data presented in the form shown in Figs.\ 7 and 8 can be done for any $\pt$ without theoretical input. The normalization for 
$\bar\xi(\phi,c)$, as defined in Eq.\ (\ref{23}), does involve a fitted parameter $\gamma$ that is not directly accessible from the data. Leaving it out, one can calculate the mean path length $\bar\ell(\phi,c)$ from geometry and the Glauber model; thus experimental  plots of $R_{AA}$ vs $\bar\ell$ can be made without changing the character of scaling. The nature of the $\bar\ell$ behavior in such a plot does depend on dynamical details at the $\pt$ considered, such as the dominance of TS or SS component in hadronization. Our general conclusion is that for pion production $R_{AA}$ behaves approximately as $e^{-2.6\bar\xi}$ with $\gamma\approx 0.1$. This is a very succinct statement of what is observed at RHIC, but can be grossly violated at LHC.

Hadron correlation in back-to-back dijets is sensitive to the path length of the associated particle on the away side. The double-hump structure observed on the two sides of $\Delta\phi=\pi$ \cite{ja, aa} has recently been examined as a function of the trigger $\phi$, revealing differences in the ``head" and ``shoulder" regions, as well as asymmetry in $\Delta\phi$ that depends on path length \cite{wh}. The interesting phenomena unfortunately cannot be compared to the findings in our study here because the trigger momentum is only in the 2-3 GeV/c range, which is much lower than what we consider. At lower $\pt$ the effect of semihard scattering is important, since such scattering is pervasive and can influence the medium in ways not accounted for by hydrodynamics. Our investigation has been on the opposite effect --- that of the medium on high $\pt$ jets. For $p_T=8$ GeV/c we find that the yield per trigger on the away side has essentially a universal $\bar\xi$ dependence for $p_b\ ^>_\sim\ 4$ GeV/c. Only the magnitude of the yield depends on $p_b$ because of the fragmentation functions, but the dependence on $\phi$ and $c$ is well described by the path-length consideration through $\bar\xi$ scaling. When expressed in terms of $D(z_T)$, the $z_T$ dependence seems to be in accord with the data. 
Verification of the $\bar\xi$ scaling in the head region at high $p_t$ and $p_b$ would give even more empirical support to our findings in dijet correlation.

At LHC that $\bar\xi$ scaling is broken when 2-jet recombination becomes important at high hard-parton density. If some aspect of our prediction is to be found valid in the data, the striking phenomenon to be seen in the preliminary result from LHC will be large $R_{AA}$ at $p_T\approx 10$ GeV/c. It would be a clear experimental indication that the conventional understanding about the effect of dense medium on hard partons (i.e., $R_{AA}<1$) is inadequate, and that the hadronization process through recombination of jet showers can obliterate the effect of momentum degradation, resulting in $R_{AA}>1$. The non-obervance of  a large value of $R_{AA}$ in excess of 1  does not falsify our consideration, since it would be due to the smallness of the overlap probability $\Gamma$ which we have no reliable way to calculate. It would reduce the impact of 2-jet recombination, but contributes to better understanding of the hadronization of partons at high density. However, if the large $R_{AA}$ is seen in the data at $\pt\sim 10$ GeV/c, it would mean that the background of the much higher-$\pt$ jets is significantly larger than what can be expected from hydrodynamical predictions by extending similar considerations from the RHIC regime.

\section*{Acknowledgment}

This work was supported  in
part,  by the U.\ S.\ Department of Energy under Grant No. DE-FG02-96ER40972 and by the National Natural Science Foundation of China under Grant No.\ 10775057, the Ministry of Education of China under Project IRT0624, and the Program of Introducing Talents of Discipline to Universities under Grant No.\ B08033.

\appendix
\section{GEOMETRICAL DETAILS}

In mapping from the initial density in the almond-shaped overlap region to the density in the initial ellipse, we have Eq.\ (\ref{19}), in which $d_1$ and $d_2$ refer to the distances from the origin to the almond and elliptical boundaries, respectively. They are defined, for any radial $\phi_r$ from the origin (not to be confused with the parton's $\phi$ from $x_0,y_0$), by
\begin{eqnarray}
({b\over 2}+d_1\cos\phi_r)^2+(d_1\sin\phi_r)^2=1,  \label{A1}  \\
({d_2\over w}\cos\phi_r)^2+({d_2\over h}\sin\phi_r)^2=1.  \label{A2}
\end{eqnarray}
The above are for $-\pi/2\le\phi_r\le\pi/2$; the case for the left half plane can be obtained by symmetry.
Solving them, we obtain
\begin{eqnarray}
{d_1\over d_2}&=&\left[-{b\over 2} \cos\phi_r+\left(1-{b^2\over 4}\sin^2\phi_r\right)^{1/2}\right]\nonumber \\
&&\times\left[{1\over w^2}\cos^2\phi_r+{1\over h^2}\sin^2\phi_r\right]^{1/2} .  \label{A3}
\end{eqnarray}
Note that $d_1/d_2=1$ when $\phi_r=0$ or $\pi/2$, as they should, since $w$ and $h$ are the $x$ and $y$ intercepts of the almond as well as the ellipse. For general $\phi_r$, $d_1/d_2$ is not far from 1, but the small difference makes the density $D(x(t),y(t))$ in Eq.\ (\ref{15}) well defined for any $(x_0,y_0)$ inside the almond and the corresponding $t_1(x_0,y_0,\phi,b)$ on the ellipse.

To determine $t_1(x_0,y_0,\phi,b)$, we substitute Eq.\ (\ref{20}) into (\ref{17}) and solve for $t_1$, getting \cite{ch}
\begin{eqnarray}
t_1 = \left[\left(B^2 + AC \right)^{1/2} - B \right]/A \ ,
\label{A4}
\end{eqnarray}
where
\begin{eqnarray}
A = \left({1\over w} \cos \phi \right)^2 +  \left({1\over h} \sin  \phi \right)^2 \ ,
\label{A5}
\end{eqnarray}
\begin{eqnarray}
B = {x_0\over w^2} \cos \phi+ {y_0\over h^2} \sin  \phi  \ , 
\label{A6}
\end{eqnarray}
\begin{eqnarray}
C = 1 - \left(x_0/w \right)^2  - \left( y_0/h\right)^2 \ .
\label{A7}
\end{eqnarray}
It then follows from Eq.\ (\ref{15}) that $\ell(x_0,y_0,\phi,b)$ can be calculated entirely from geometry and therefore $\bar\ell(\phi,b)$ as well, using Eq.\ (\ref{23}) without the multiplicative factor $\gamma$.

\end{document}